\documentclass[10pt]{article}

\usepackage[final]{epsfig}
\usepackage{amsmath}
\usepackage{parskip}
\usepackage{cite}

\def\lsim{\mathrel{\rlap{\lower4pt\hbox{\hskip1pt$\sim$}}
    \raise1pt\hbox{$<$}}}         
\def\gsim{\mathrel{\rlap{\lower4pt\hbox{\hskip1pt$\sim$}}
    \raise1pt\hbox{$>$}}}         

\def\overleftrightarrow#1{\vbox{\ialign{##\crcr
    $\leftrightarrow$\crcr
    \noalign{\kern 1pt\nointerlineskip}
    $\hfil\displaystyle{#1}\hfil$\crcr}}}

\begin{document}

\hspace{12cm}{\bf DUKE-TH-04-262}\\

\begin{center}
{\bf A comprehensive description of multiple observables in heavy-ion collisions at SPS}

\end{center}
\begin{center}
{Thorsten Renk$^{a}$}

{\small \em $^{a}$ Department of Physics, Duke University,
PO Box 90305,  Durham, NC 27708 , USA}

\end{center}
\vspace{0.25 in}

\begin{abstract}
Combining and expanding on work from previous publications, a model for the evolution of
ultrarelativistic heavy-ion collisions at the CERN SPS for 158 AGeV beam energy is presented.
Based on the assumption of thermalization and a parametrization of the space-time
expansion of the produced matter, this model is able to describe a large set of observables
including hadronic momentum spectra, correlations and abundancies, the emission of
real photons, dilepton radiation and the suppression pattern of charmonia. Each of these
obervables provides unique capabilities to study the reaction dynamics and taken together
they form a strong and consistent picture of the evolving system. Based on the emission
of hard photons, we argue that a strongly interacting, hot and dense system with temperatures
above 250 MeV has to be created early in the reaction. Such a system is bound to be
different from hadronic matter and likely to be a 
quark-gluon plasma, and we find that this assumption is in line with the subsequent
evolution of the system that is reflected in other observables.

\end{abstract}

\vspace {0.25 in}

\section{Introduction}
\label{sec_introduction}

Lattice simulations (e.g. \cite{lat1}) predict that  QCD
undergoes a phase transformation at a temperature $T_C \approx 150 - 170$ 
MeV \cite{lat2,lat3} from a confined hadronic phase to a
phase, the quark-gluon plasma (QGP), in which quarks and gluons constitute the relevant degrees of freedom
and the chiral condensate vanishes. 
Experimentally, this prediction can only be tested in ultrarelativistic
heavy-ion collisions. However, finding evidence, and ultimately proof, for the 
creation of a quark-gluon plasma faces several difficulties. Arguably the greatest challenge
is to link experimental observables to quantities measured on the lattice.

For a large set of observables, the evolution of the expanding medium is a key
ingredient for their theoretical description. It is also the place where results
from lattice QCD fit in: assuming a thermalized system is created,
its evolution is governed by the equation of state (EoS). 
Hence, in order to test the lattice QCD predictions, one has to start with this assumption
and show that it leads to
a good description of the experimental data.

It is the purpose of the present paper to show that such a description can be
achieved in the case of 158 AGeV Pb-Pb and Pb-Au collisions at the CERN SPS. 
We summarize and expand on the results of \cite{Dileptons, Thesis,
Photons, Hadrochemistry, Charm, Amruta, HBT} by developing
a framework based on the EoS determined in lattice QCD.
Using eikonal calculations to constrain the initial state and measured bulk hadronic
properties to establish the final state, we parametrize the
evolution in between in a way that is motivated by hydrodynamical
calculations. We then  use this parametrization to calculate other observables not
connected to the bulk hadronic matter
such as dilepton and photon emission and charmonium suppression. 
In all cases, we find good agreement with the data with the
same set of model parameters.
We explicitly discuss how each observable reflects underlying scales of the 
evolution, to what degree information from a particular observable constrains
alternative evolution scenarios and what we can learn about the relevant degrees
of freedom in the medium.

\section{The fireball model framework}
\label{sec_framework}

\subsection{Expansion and flow}

We do not aim at a microscopical description of the expansion but instead at an effective
parametrization based on available data. In order to simplify computations, we introduce several 
assumptions which lead to an idealized picture of the expansion process. The aim of the
model is to demonstrate that there is one set of scales characterizing the expansion
that is in reasonable agreement with a large set of observables, not to provide a detailed 
description of every observable.

Our fundamental assumption is that an equilibrated system is formed a short proper time
$\tau_0$ after the onset of the collision, which subsequently expands isentropically.
As soon as the mean free path of particles in the medium exceeds its dimensions,
kinetic decoupling occurs at a proper time $\tau_f$. The matter then ceases to
be in thermal equilibrium and we assume that no significant interactions occur later.  

During the expansion phase, we posit that the entropy density of the system can
be described by 
\begin{equation}
s(\tau, \eta_s, r) = N \overline{R}(r,\tau) \cdot \overline{H}(\eta_s, \tau)
\end{equation}
with $\tau $ the co-moving proper time of
a given volume element and $\eta_s = \frac{1}{2}\ln (\frac{t+z}{t-z})$ the spacetime
rapidity. $\overline{R}(r, \tau)$ and $\overline{H}(\eta_s, \tau)$ are two functions 
describing the evolving shape of the distribution,
and $N$ is a normalization factor chosen such that the total entropy is 
$S_0 = \int d^3 r s(\tau, \eta_s, r)$. In the above expressions we have neglected
angular asymmetries for collisions with a finite impact parameter, we expect their
influence to be small as long as we do not  discuss very peripheral collisions or observables,
such as elliptic flow, with an explicit angular dependence.

In oder to simplify our calculations, we choose $\overline{R}(r, \tau)$ and 
$\overline{H}(\eta_s, \tau)$
as box profiles $\overline{R}(r, \tau) = \theta(R(\tau)-r)$, 
$\overline{H}(\eta_s, \tau) = \theta(H_s(\tau) - \eta_s) \theta(\eta_s) + \theta(\eta_s -H_s(\tau)) \theta(-\eta_s)$. For this choice, thermodynamical
parameters become functions of proper time $\tau$ only. We will later argue that 
this simplified choice of the distribution function gives a fair representation of the
physics at midrapidity, where most observables are measured, but fails to reproduce the conditions
at forward rapidities. For bulk observables, like particle numbers, obtained by integrating
over the whole volume, the detailed choice is irrelevant or reduced to a second order effect
due to detector acceptance.

Thus, the expansion is governed by the scale parameters $R(\tau)$ and $H(\tau)$. Their
growth with $\tau$ is a consequence of collective flow of the thermalized matter.
For transverse flow we assume a linear relation between radius $r$ and
transverse rapidity $\rho = \text{atanh } v_\perp(\tau) =  r/R_c(\tau) \cdot \rho_c(\tau)$
with $\rho_c(\tau) = \text{atanh } a_\perp \tau$. The longitudinal flow profile 
is dictated by the requirement that an initially homogeneous distribution of matter remains
homogeneous at all times.

We start with the experimentally measured width of the rapidity 
interval of observed hadrons $2\eta_f^\text{front}$ at breakup. From this, we compute the longitudinal velocity of the fireball front at kinetic freeze-out $v_f^\text{front}$.
We do not require the initial expansion velocity $v_0^\text{front}$ to coincide
with $v_f^\text{front}$ but instead allow for a longitudinally accelerated expansion. 
This implies that during the evolution $\eta = \eta_s$ 
where $\eta$ is the momentum rapidity,
is not valid .
Initially, we characterize all possible trajectories of volume elements
by a parameter $c$ such that $v_0^z(c) = cv_0^\text{front}$, $c \in [-1,1]$; hence we 
start with an expansion that is boost-invariant over the rapidity
interval $-\text{atanh } v_0^\text{front} \leq \eta \leq \text{atanh } v_0^\text{front}$.
For the final state, we require $\text{atanh }v_f^\text{front} = \eta_f^\text{front}$, i.e.
the distribution has to fill the experimentally observed rapidity interval.

\begin{figure}[htb]
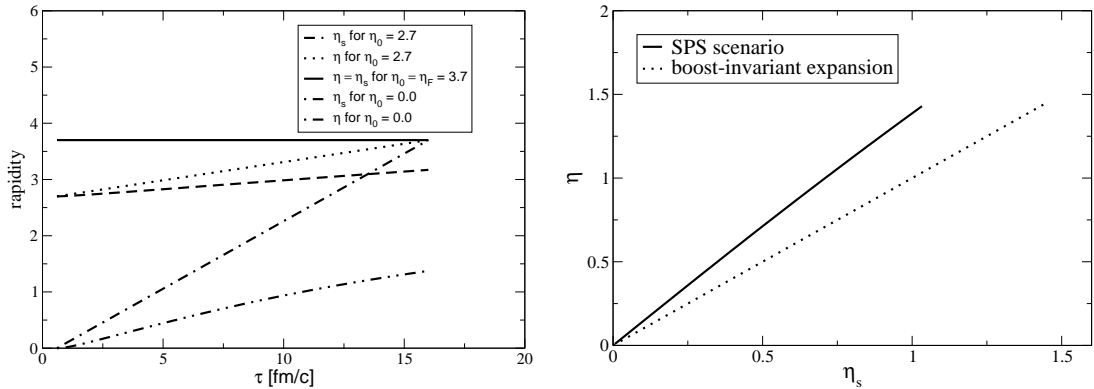

\begin{center}
\vspace*{3ex}
\epsfig{file=rapidity.eps, width=7.0cm} \hspace{0.2cm}
\epsfig{file=eta_of_eta_s_sps.eps, width=7.0cm}
\end{center}
\caption{\label{F-Rapidities}Left: Different scenarios for the longitudinal 
fireball expansion, all tuned to produce the same final rapidity distribution. 
Shown are rapidity $\eta$ and spacetime rapidity $\eta_s$ for a standard boost-invariant 
scenario without long. acceleration (solid), for complete initial stopping and re-expansion 
($\eta$: dash-dotted $\eta_s$: dash-dot-dotted) and and intermediate scenario ($\eta$: dotted
$\eta_s$: dashed). 
Right: $\eta$ as a function of $\eta_s$ at freeze-out time $\tau_f$ for the
SPS scenario described in this paper.  }
\end{figure}

In order to match initial and final state, an acceleration term has to be introduced.
We assume that the acceleration is driven by the local conditions
in and around a volume element (as it would be in hydrodynamic calculations),
hence it has to be a function of proper time.
To maintain a spatially homogeneous distribution of matter at all
times, the acceleration furthermore must correspond to a re-scaling of
the velocity field. Therefore, if $a_z(\tau)$ is the acceleration acting
on the fireball front, the trajectory characterized by $c$ feels the
acceleration $c a_z(\tau)$.  This implies
\begin{equation}
\label{E-tau}
\tau = \int_{t_0}^{t} \sqrt{1 - v^z(c, \tau(t', z(t')))^2} dt'
\qquad \text{with} \qquad
v^z(c, \tau) = \frac{c \cdot v_0^z + c\int_{\tau_0}^\tau a_z(\tau') d\tau'}{1 + v^z_0 c^2\int_{\tau_0}^\tau a_z(\tau') d\tau'} .
\end{equation}
During the evolution, the volume element moves the distance
\begin{equation}
z(c,t) = \int_0^{t} v^z(c,\tau(t, z(t)))  dt
\end{equation}
in the c.m. frame. We solve the set of equations numerically for all values of $c$ by integrating
the trajectory forward in time until a fixed $\tau$ in Eq.~(\ref{E-tau}) is reached. 
The resulting pairs ($t(c), z(c)$) define a curve
of constant proper evolution time. We can compute the length of this curve using the inavriant volume formula $V = \int d \sigma_\mu u^\mu$ with $\sigma_\mu$ an element of the
freeze-out hypersurface and $u^\mu$ the four-velocity of matter. To good accuracy,
volume elements lie on the curve $\tau = const. = \sqrt{t(c)^2-z(c)^2}$ 
and the flow pattern can be approximated
by a linear relationship between rapidity $\eta$ and  spacetime rapidity $\eta_s$ as
 $\eta(\eta_s) = \eta^\text{front}\eta_s/\eta_s^\text{front}$
where $\eta_s^\text{front}$ is computed on the $c=1$ trajectory. In this case, the longitudinal
extension in the proper volume formula yields
\begin{equation}
L(\tau) \approx 2 \tau \frac{\text{sinh }(\zeta -1) \eta_s^\text{front}}{(\zeta -1)} 
\end{equation}
with $\eta_s^\text{front}(\tau)$
the spacetime rapidity
of the accelerating cylinder front and $\zeta(\tau) = \eta^\text{front}(\tau)/\eta_s^\text{front}(\tau)$ the mismatch between
spacetime rapidity and rapidity introduced by the accelerated motion. This is an approximate
generalization of the boost-invariant relation $L(\tau) = 2 \eta^\text{front} \tau$ which can be derived
for non-accelerated motion. The volume of the fireball at given proper time can then
be found as $V= \pi R(\tau)^2 L(\tau)$ where we have neglected relativistic effects in the
transverse dynamics (we have estimated that those are on the order of 10\% at maximum for the volume
and thus well within the parametric uncertainties).
We illustrate these concepts in Fig.~\ref{F-Rapidities}

\subsection{Parameters of the expansion}

We now address the proper time dependence of the two
scales $R(\tau)$ and $H(\tau)$.
The parameters entering $v_z^\text{front}(\tau)$ 
(cf. Eq.~(\ref{E-tau})) are determined using the ansatz
$a_z = c_z \cdot p(\tau)/\epsilon(\tau)$
which allows a soft point in the EoS where the ratio $p/\epsilon$ gets small
to influence the acceleration pattern. $c_z$ and $v_0^\text{front}$ in (\ref{E-tau}) are then
model parameters governing the longitudinal expansion and fitted to data.
For the radial expansion we use
\begin{equation}
R(\tau) = R_0 + c_T \int_{\tau_0}^\tau d \tau'  \int_{\tau_0}^{\tau'} d \tau'' \frac{p(\tau'')}{\epsilon(\tau'')}
\end{equation}
The initial radius $R_0$ is taken from overlap calculations. This leaves a parameter
$c_T$ determining the strength of the transverse acceleration which is also fitted to
data. The final parameter characterizing the expansion is its endpoint given by $\tau_f$,
the breakup proper time of the system. It is determined by the condition that the temperature
$T(\tau)$ drops below the freeze-out temperature $T_F$. 

\subsection{Thermodynamics}

We calculate the total entropy $S_0$ by fixing the entropy per baryon from the number
of produced particles per unit rapidity and the number of participant
baryons \cite{Dileptons,ENTROPY-BARYON}. Assuming isentropic expansion
the entropy density $s$ at a given proper time
is then determined by $s=S_0/V(\tau)$.

We describe the EoS in the partonic phase by a quasiparticle
interpretation of lattice data which has been shown to reproduce lattice
results both at vanishing baryochemical potential $\mu_B$ \cite{QP1} and
finite $\mu_B$ \cite{QP2} and is able to extrapolate the lattice results
obtained with quark masses $m\sim T$ down to physical masses.

For a thermalized hot hadronic system near the phase transition, 
lattice computations of the EoS offer little guidance 
since the quark masses and hence the hadron masses are still unphysically large.
On the other hand, many of the
hadronic states relevant close to the phase transition are
poorly known and a first principle calculation of the thermodynamics of an
interacting hadron gas near this point faces serious difficulties.
We circumvent the problem by calculating thermodynamic properties of a hadron
gas at kinetic decoupling where interactions cease to be important and an ideal gas
is presumably a valid description. Determining
the EoS at this point using the statistical hadronization described in
section \ref{S-Hadrochemistry}, we choose a smooth interpolation between decoupling temperature
$T_F$ and transition temperature $T_C$ to the EoS
obtained in the quasiparticle description. 

Using the EoS and $s(\tau)$, we compute
the parameters $p(\tau), \epsilon(\tau)$ and $T(\tau)$.
Since the ratio $p(\tau)/\epsilon(\tau)$ appears in the expansion
parametrization, we solve the model self-consistently by iteration.

\subsection{Solving the model}

Several of the model parameters are fixed by measurements,
such as the total entropy $S_0$  or the final rapidity interval
$\eta_f^\text{front}$. Others are calculated, such as 
the initial transverse overlap radius $R_0$ or the number of participant nucleons $N_\text{part}$
which are  obtained by eikonal calculations. 
Only the initial rapidity interval $\eta^\text{front}(\tau_0)$ (via $c_z$), the final transverse
velocity $v_\perp(\tau_f)$ (via $c_T$) and the freeze-out temperature $T_F$ are fit
parameters and adjusted to the hadronic momentum spectra and HBT correlations. 
Thus, total multiplicity is governed
by the input $S_0$, the relative multiplicity of different particle species
is calculated in a statistical hadronization framework (see section \ref{S-Hadrochemistry})
whereas the shape of the transverse mass spectra and the HBT radii determine
the transverse and longitudinal expansion parameters and the freeze-out 
temperature.
The equilibration time $\tau_0$ cannot be obtained from fits to hadronic
observables which reflect the final state; for the time being we choose
$\tau_0 = 1$ fm/c and discuss variations of this parameter later.

\subsection{Particle emission and HBT}

We calculate particle emission throughout the whole lifetime of the fireball
by  evaluating the Cooper-Frye
formula
\begin{equation}
E \frac{d^3N}{d^3p} =\frac{g}{(2\pi)^3} \int d\sigma_\mu p^\mu
\exp\left[\frac{p^\mu u_\mu - \mu_i}{T_f}\right] = d^4 x S(x,p)
\end{equation}
with $p^\mu$ the momentum of the emitted particle and $g$ its 
degeneracy factor from the fireball surface. For emission
hypersurfaces with a spacelike normal we include a $\theta$-function
preventing emission from the surface back into the system.
Note that the factor $d\sigma_\mu p^\mu$ contains the spacetime rapidity
$\eta_s$ and the factor $p^\mu u_\mu$ the rapidity $\eta$. Since these are
in general not the same in our model, the analytic expressions
valid for a boost-invariant scenario \cite{ThermalPhenomenology} 
do not apply. 

Pionic HBT correlation radii are obtained using the common
Cartesian parametrization for spin 0 particles
\begin{equation}
C(q,K) -1 = \exp \left[ -q_\text{o}^2R_\text{out}^2(K) - q_\text{s}^2R_\text{side}^2(K)-
q_\text{l}^2R_\text{long}^2(K)-2q_\text{o}q_\text{l}R^2_\text{ol}(K)\right]
\end{equation}
(see e.g. \cite{HBTReport, HBTBoris} for
an overview and further references)
for the correlator.
Here, $K = \frac{1}{2} (p_1 + p_2)$ is the averaged momentum of the correlated pair 
with individual momenta $p_1,p_2$ and $q = (p_1 - p_2)$ the relative momentum.
The transverse correlation radii $R_{\text{out,side, long}}$ follow from the emission function
as
\begin{equation}
\label{ERSideBase}
R_\text{side}^2 = \langle \tilde{y}^2 \rangle \qquad R_\text{out}^2 = \langle (\tilde{x} - \beta_\perp\tilde{t})^2 \rangle \qquad R_\text{long}^2 = \langle \tilde{z}^2 \rangle
\end{equation}
with $\beta_\perp$ the transverse velocity of the emitted pair, 
$\tilde x_\mu = x_\mu - \langle x_\mu \rangle$ and
\begin{equation}
\langle f(x) \rangle(K) = \frac{\int d^4 x f(x) S(x,K)}{\int d^4 x S(x,K)}
\end{equation}
an average with the emission function.

\subsection{Fit results --- $m_t$-spectra and HBT radii}

\begin{figure}[htb]
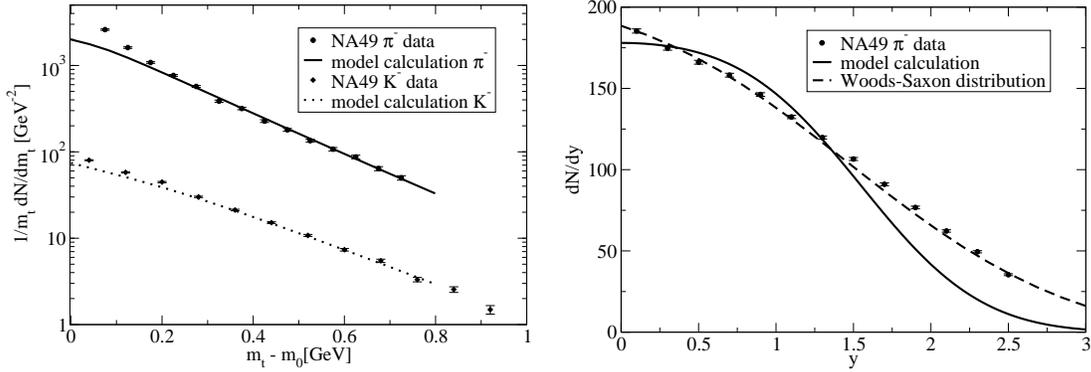

\begin{center}
\vspace*{0.5cm}
\epsfig{file=m_t-sps.eps, width=7.0cm} \hspace{0.2cm}
\epsfig{file=dNdy-sps.eps, width=7.0cm}
\end{center}
\caption{\label{F-Hadronspectra}Left panel: Transverse mass spectra of $\pi^-$ and
K$^-$ measured by NA49 \cite{NA49-1} compared with the model results. Right panel:
The rapidity distribution of $\pi^-$ measured by NA49 \cite{NA49-1} (defining
0 as the rapidity of the center of mass system) compared with the model result
(solid) and a calculation replacing the box density profile by a Woods-Saxon
distribution with skin thickness $\Delta y =0.37$ (dashed).  }
\end{figure}

The model is fitted to the recent SPS data by NA49 \cite{NA49-1}, 
hence there are small differences to the results of \cite{FREEZE-OUT}
used in \cite{Dileptons, Thesis ,Photons, Charm, Amruta}
As an example, we show the resulting transverse mass spectra for $\pi^-$ and $K^-$
in Fig.~\ref{F-Hadronspectra} (left) and the rapidity distribution for
$\pi^-$ (right) compared with the NA49 data.

There is a contribution to the $\pi^-$ transverse mass spectrum 
coming from the vacuum decay of resonances after
$\tau_f$. The model is able to calculate the magnitude
of this contribution using statistical hadronization but not
its $m_t$ distribution, therefore it is left out when we compare with data.
This explains the mismatch at low transverse mass and that the
integrated spectrum does not yield the multiplicity indicated by the data.
The same is visible in the K$^-$ spectrum, albeit less pronounced.
Apart from this aspect, the model describes the transverse mass spectra well.

The fact that particles inside the fireball volume are characterized by a thermal momentum distribution
leads to a smearing of the box-profile in rapidity when particle emission is calculated,
therefore the resulting $\pi^-$ distribution in rapidity does not exhibit
a sharp cutoff. The deviation from the data at forward rapidities is 
caused by the simplifying assumption of a homogeneous rapidity density, it is not a
failure of the framework as such and indicates that this approximation should
not be used for $y >1.5$. We illustrate this by replacing the
box density distribtion in rapidity  with a Woods-Saxon distribution
with a skin thickness parameter $\Delta y =0.37$ which dramatically improves the 
agreement with the
data. We have checked that this choice does not significantly alter the transverse
mass spectra at $y=0$.

\begin{figure}[htb]
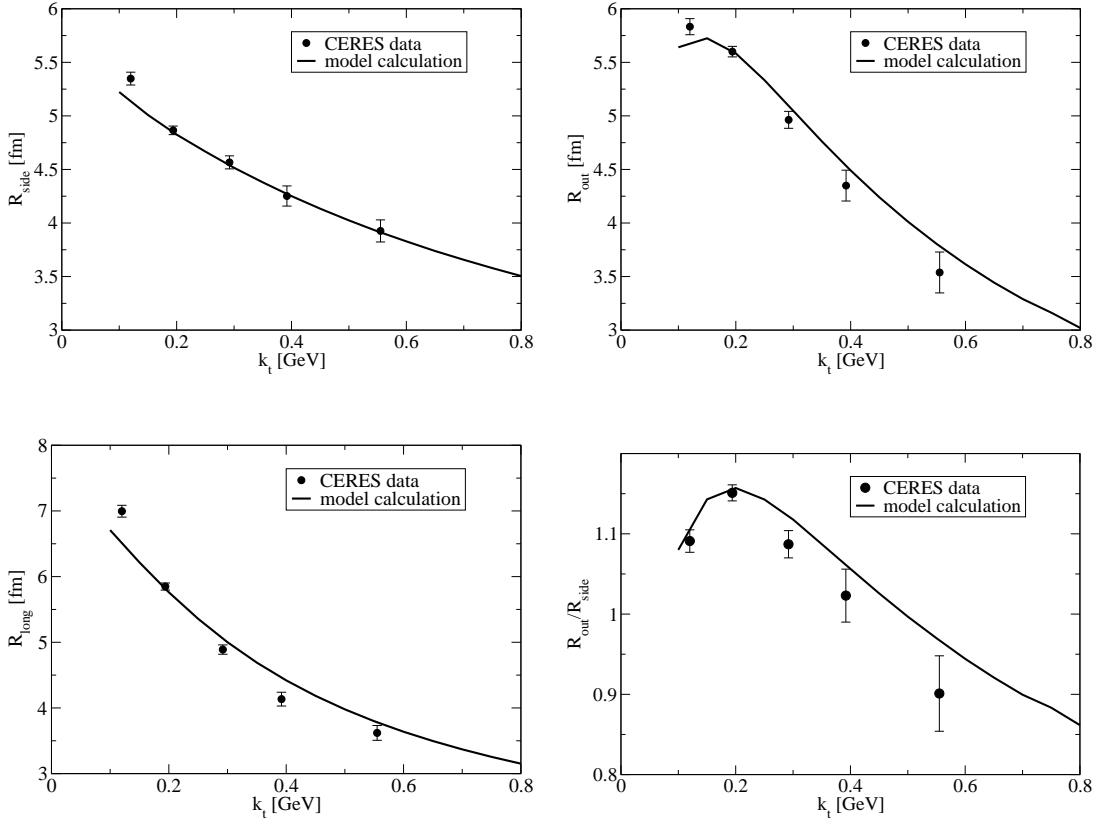

\begin{center}
\vspace*{0.5cm}
\epsfig{file=R_side.eps, width=7.0cm} \hspace{0.2cm}
\epsfig{file=R_out.eps, width=7.0cm} 

\vspace*{0.8cm}
\epsfig{file=R_long.eps, width=7.0cm} \hspace{0.2cm}
\epsfig{file=R_ratio.eps, width=7.0cm}
\end{center}
\caption{\label{F-HBT} HBT Correlation radii obtained by the CERES collaboration
\cite{CERES-HBT} as a function of transverse pair momentum $k_t$ 
compared with the model results. No systematic errors are included
in the experimental errorbars.}
\end{figure}

The resulting HBT correlation parameters are shown in Fig.~\ref{F-HBT}.
In general, the agreement with the data is good, however the low $k_t$ region seems
to be underestimated. Note however that our calculation does not contain
the model independent correction terms which tend to increase the correlation radii
for small momenta (see our results in \cite{HBT}) and that the 
experimental data do not include systematic errors.

\subsection{The fireball evolution}

\begin{figure}[htb]
\begin{center}
\epsfig{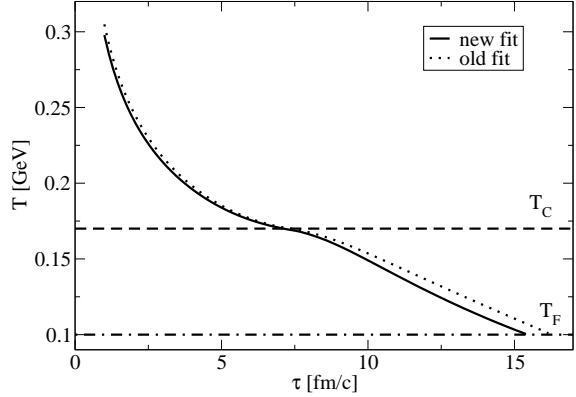}
\end{center}
\caption{\label{F-TEvolution}Temperature evolution of the fireball for SPS 158 AGeV
5\% centeral Pb-Pb collisions, shown are the old fit used in \cite{Dileptons, Thesis, Photons, Charm, Amruta} and the scenario based on recent data presented in this work.}
\end{figure}

The model for 5\% central 158 AGeV Pb-Pb collisions at SPS is characterized by
the following scales: Initial longitudinal velocity $v_0^\text{front} = 0.54c$, equilibration
time $\tau_0 = 1$ fm/c, 
initial temperature $T_0 = 300$ MeV, duration of the QGP phase $\tau_\text{QGP} = 6.5$ fm/c,
duration of the hadronic phase $\tau_\text{had} = 8.5$ fm/c, total lifetime $\tau_f - \tau_0
= 15$ fm/c,  r.m.s radius at freeze-out $R_f^\text{rms} = 8.55$ fm, transverse expansion
velocity $v_{\perp f} = 0.57 c$ and freeze-out temperature $T_F= 100$ MeV . Fig.~\ref{F-TEvolution} shows the resulting temperature
evolution in proper time and compares it with the previous fit results based on older data 
used in \cite{Dileptons,
Thesis, Photons, Charm, Amruta}.

The difference between the two fits is not large --- the new data require a slightly
larger $v_0^\text{front}$, leading to a reduced initial temperature, but this change
is well within the model uncertainties. A more pronounced difference shows up in the late
hadronic evolution where a larger $v_\perp^f$ is required, leading to faster cooling
and a reduced lifetime in the new fit. Surprisingly, this amounts to an increase of the four-volume
of the hadronic phase of about 5\%, the shorter time duration is slightly outweighed 
by a faster volume increase.

Based only on a fit to hadronic observables, we thus find a system which is characterized by a
sizeable initial compression of matter with subsequent re-expansion 
($v_0^\text{front} < v_f^\text{front}$).
This has several important consequences: First, any estimate of the initial temperature or
total lifetime based on a boost-invariant expansion such as the well-known relation
$
R_\text{long} = \tau_f(T_F/m_t)^{1/2}
$
connecting $R_\text{long}$ with the breakup time \cite{Sinyukov}
does not apply. Instead, a system undergoing accelerated longitudinal expansion is initially
more strongly compressed (implying a larger initial energy density and temperature)
and takes more time to expand to a given volume while arriving at the observed longitudinal flow.
Hence, the four-volume of such a system is also larger than in the standard boost-invariant
expansion scenario. We argue later that the difference between the two cases is
experimentally accessible by measuring electromagnetic probes.

For the discussion of some observables, we require the fireball evolution for other
than 5\% central collisions. In this case, we scale the total entropy with the number
of collision participants and reduce the initial radius of the system to reproduce
the smaller overlap area of the nuclei while neglecting the angular asymmetry.
Keeping the parameter $c_\perp$ at its value, the reduced entropy
leads to an earlier freeze-out and hence naturally reduces transverse flow.
For the initial longitudinal motion, we linearly interpolate between the value
found for central collisions and the observed rapidity loss in p-p collisions
(where no re-expansion occurs) in the number of collision participants and
refit $c_z$ accordingly.

\section{Statistical hadronization at the phase boundary}

\label{S-Hadrochemistry}

In order to connect the fireball entropy with the measured particle spectra,
we have to specify how a hadronic system at the phase boundary is created. 
Statistical models
are extremely successful in describing the measured ratios of different hadron
species for a range of collision energies from SIS to RHIC
(see e.g. \cite{PBM1,PBM2, PBM3, Broniowski}).

The basic assumption of statistical hadronization is that the particle content of the fireball 
can be found by considering 
a system of (non-interacting) hadrons in chemical equilibrium at $T_C$,
described by the grand canonical ensemble, which subsequently undergoes decay
processes. This ensemble is characterized by the temperature $T_C$, the 
baryochemical potential $\mu_B$ and the strange potential $\mu_s$.
In the present model framework, we can calculate all these parameters from
the fireball evolution and obtain hadron ratios parameter-free.

Employing the grand canonical ensemble, 
the density for each particle species $n_i$ is given by
\begin{equation}
\label{E-GCE}
n_i = \frac{d_i}{2\pi^2}\int_0^\infty
\frac{p^2 dp}{\exp\{[E_i(p)-\mu_i]/T_C\}\pm 1}.
\end{equation}
Here, $d_i$ denotes the degeneracy factor of particle 
species $i$ (spin, isospin, particle / antiparticle), 
the +(-) sign is used for fermions (bosons) and 
$E_i(p)=\sqrt{m_i^2 + p^2}$. $m_i$ stands for the
particle's vacuum mass.
We use a value of 170 MeV for the critical temperature
$T_C$ as determined in lattice calculations
for two light and one heavy flavours \cite{lat2, lat3}.
The chemical potential $\mu_i = \mu_B B_i - \mu_S S_i$ takes care of conserved
baryon number $B_i$ and strangeness $S_i$ for each species and we neglect 
a (small) contribution $-\mu_{I_3} I_i^3$
coming from the isospin asymmetry. 
The conservation of the number of 
participant baryons $N_\text{part}$ and the requirement of 
vanishing net strangeness uniquely determines $\mu_B$ and
$\mu_S$ for a given volume $V = V(T_C)$. This volume
can be calculated as $V(T_C) = S_{tot}/s(T_C)$
with $s(T_C)$ determined by the EoS in the partonic phase,
thus linking the model calculation again with lattice results.

We include all mesons and mesonic resonances up to masses of
1.5 GeV and all baryons and baryonic resonances up to masses
of 2 GeV in the model and calculate
their decay into particles which are stable or long-lived as compared
to the fireball, such as $\pi, K, \eta, N, \Lambda, \Sigma$ and
$\Omega$. In order to account for repulsive interactions between particles at small distances 
we assume a hard core radius
$R_{C}$ of 0.3 fm (based on results for p-p collisions \cite{HardCore})
for all particles and resonances and correct for
the excluded volume self-consistently. Particle and resonance
properties are taken from \cite{ParticleDataBook}.
For resonances with large width,
we integrate Eq.~(\ref{E-GCE}) over the mass
range of the resonance using a Breit-Wigner distribution.

\begin{figure}[htb]
\begin{center}
\epsfig{file=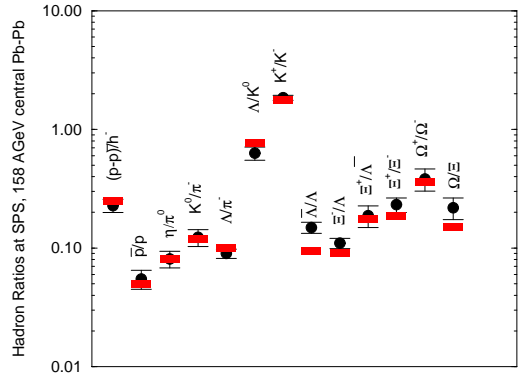, width=7.5cm}
\end{center}
\caption{\label{F-RatiosStd}Hadron ratios in the statistical hadronization
model (dashed bands) as compared to experimental results (filled circles) for SPS,
158 AGeV central Pb-Pb collisions \cite{R1,R2,R3,R4,R5,R6,R7,R8,R9,R10,R10a}.}
\end{figure}

The result (Fig.~\ref{F-RatiosStd}) is in good agreement with the experimental data
\cite{R1,R2,R3,R4,R5,R6,R7,R8,R9,R10,R10a}.
The ratios are (for given $T_C$) determined
by the volume which in turn determines $\mu_B$.
We may infer from this agreement that the value of $T_C$
as measured on the lattice is, in combination with the statistical hadronization
approach, compatible with the data and that the EoS, via $s(T_C)$, 
leads to a reasonable volume at the phase transition. Therefore,
hadron ratios are insensitive to the spacetime expansion parametrized in the model but
test mainly properties of the EoS.

The main impact of the results on the fireball evolution is via the EoS:
Decay processes of high-lying resonances created at $T_C$ lead
to on overpopulation of pion phase space during the evolution in the hadronic
phase when strong decay processes are still in equilibrium with their back reaction
but weak decays are not. This effect has a crucial influence on the EoS and we incorporate it by 
allowing for a finite pion chemical
potential $\mu_\pi (\tau)$ that rises to good approximation linearly in proper time from 0 at $\tau=\tau_C$ 
to 110 MeV at $\tau = \tau_f$ (cf. \cite{Goity}).

\section{Electromagnetic observables}

\subsection{Dilepton emission}

The lepton pair emission rate from a hot domain populated by particles in
thermal equilibrium at temperature $T$ is proportional to the imaginary part
of the spin-averaged photon self-energy, with these particles as
intermediate states. The thermally excited particles annihilate to yield a
time-like virtual photon with four-momentum $q$ which decays subsequently into
a lepton-antilepton pair. The differential pair production rate is given by \cite{Wong}
\begin{equation}
\frac{dN}{d^4 x d^4q}  =  \frac{\alpha^2}{\pi^3 q^2} \ \frac{1}{e^{\beta q^\mu u_\mu}
- 1} \ \mbox{Im}\bar{\Pi}(q, T) = \frac{\alpha^2}{12\pi^4} \frac{R(q, T)}{e^{\beta q^\mu u_\mu}
- 1}  ,\label{dileptonrates}
\end{equation}
where $\alpha = e^2/4\pi$, $\beta = 1/T$, $u_\mu$ the four-velocity of the emitting volume
element and we have neglected the lepton
masses. We have defined $\bar{\Pi}(q) = -\Pi^\mu_{\ \mu} /3$ and introduce the averaged photon spectral function $R(q, T) = (12\pi/q^2) \ \mbox{Im}\bar{\Pi}(q,T)$. Here $\Pi^\mu_{\
\mu}$ denotes the trace over the thermal photon self-energy which is
equivalent to the thermal current-current correlation function
\begin{equation}
\Pi_{\mu\nu}(q,T) = i \int d^4 x \ e^{iqx} \langle \mathcal{T} j_{\mu}(x)
j_{\nu}(0) \rangle_\beta,
\end{equation}
where $j_\mu$ is the electromagnetic current.
Eq.(\ref{dileptonrates}) is valid to leading order in the electromagnetic
interaction and to all orders in the strong interaction.
The information about the strong interaction dynamics is now encoded in the spectral
function. For its computation, we have to use different techniques for dealing with
partonic and hadronic degrees of freedom.

\subsubsection{Spectral function in the QGP phase}

\label{subsec_qgp}

As long as the thermodynamically active degrees of freedom are quarks and
gluons, the timelike photon couples to the continuum of thermally
excited $q\overline{q}$ states and subsequently converts into a charged lepton pair.
The calculation of the photon spectral function  at the one-loop level is performed using
standard thermal field theory methods. The well-known leading-order result
for bare quarks and gluons as degrees of freedom is \cite{Wong}:
\begin{equation}
\label{E-ImPi}
\begin{split}
\text{Im}\overline{\Pi}(q^0, {\bf q},T) &= -\frac{q^2}{12\pi} \cdot 3 \sum_{f=u,d,s} \theta(q^2-4m_f^2) e_f^2
\left(1+\frac{2m_f^2}{q^2}\right) \gamma(q^2)\\
\times&\left(1+2\left[\frac{T}{|{\bf q}|} \frac{1}{\gamma(q^2) }
\ln \left(\frac{f_D\left(\frac{q_0}{2} -\frac{|{\bf q}|}{2} \gamma(q^2) \right)}
{f_D\left(\frac{q_0}{2} +\frac{|{\bf q}|}{2} \gamma(q^2) \right)}\right) -1\right]
\right),
\end{split}
\end{equation}
where $q = (q^0, {\bf q})$ is the four-momentum of the virtual photon, $e_f$ the quark electric
charge, $\gamma(q^2) = \sqrt{1-\frac{4m_f^2}{q^2}}$ 
and $m_f$ the quark mass of flavour $f$. This result, however, is  modified by 
perturbative corrections in $\alpha_s$ that take into account
interactions in the plasma.

We compute these within our quasiparticle description \cite{QP1, QP2}. Here, the degrees of 
freedom in the partonic phase acquire an effective temperature dependent mass $m(T)$. 
Furthermore, the effect of confinement near the phase transition is parametrized by
a reduction factor $C(T)$ multiplying the distribution functions (see
references for details). 

It is straightforward to insert thermal masses instead of bare quark masses into Eq.~\ref{E-ImPi}.
$C(T)$ is equivalent to a fugacity factor, it describes the reduced occupation of available
states for quarks and antiquarks in the medium due to confinement. Since Eq.~\ref{E-ImPi} describes
the annihilation of a $q\overline{q}$ pair into a virtual photon,
the probability of finding a quark and antiquark in the medium is reduced by a factor $C(T)^2$ 
in this approximation.
 
\subsubsection{Spectral function in the hadronic phase}

Below $T_C$, confinement sets in and the effective degrees of freedom
change to colour singlet, bound $q\bar{q}$ or $qqq$
($\bar{q}\bar{q}\bar{q}$) states. The photon couples now to the lowest-lying
dipole excitations of the vacuum, the hadronic $J^P = 1^-$ states: the
$\rho$, $\omega$ and $\phi$ mesons and multi-pion states carrying these  same
quantum numbers.
There is a considerable uncertainty in the calculation of properties of
hadronic matter near the phase boundary. In order to minimize the theoretical
uncertainty, we compare several different approaches:

The first approach (referred to as 'chiral model' in the following) is
based on \cite{SW00, SW01,KKW2}:
The electromagnetic current-current correlation function can
be computed from an effective
Lagrangian which approximates the $SU(3)$ flavour sector of QCD at low
energies. An appropriate model for our purposes is the {\em Improved Vector
Meson  Dominance} model combined with chiral dynamics of pions and kaons as
described in \cite{KKW1}. Within  this model, the following relation between
the imaginary part of the irreducible photon self-energy  $\mbox{Im}
\bar{\Pi}$ and the vector meson self-energies $\Pi_V(q)$ in vacuum is derived:
\begin{equation}
\mbox{Im} \bar{\Pi}(q) = \sum \limits_V \frac{\mbox{Im}
\Pi_V(q)}{g_V^2} \ |F_V(q)|^2, \label{ImBarPi}  \quad F_V(q) = \frac{\left( 1-
g/g^0_{V} \right)q^2 - m_V^2}{q^2 - m_V^2 + i  \mbox{Im}\Pi_V(q)},
\end{equation}
where $m_V$ are the (renormalized) vector meson masses,
$g^0_V$ is the $\gamma V$ coupling and $g$ is the vector meson coupling
to the pseudoscalar Goldstone bosons $\pi^\pm, \pi^0$
and $K^\pm, K^0$. Eq.(\ref{ImBarPi}) is valid for a virtual photon with
vanishing three-momentum $\mathbf{q}$. For finite three-momenta there exist two
scalar functions $\bar{\Pi}_L$ and $\bar{\Pi}_T$, because the existence of a
preferred frame of reference (the heat bath) breaks Lorentz invariance, and
one has to properly average over them. However, for simplicity we
approximate the problem by taking the limit  $|\mathbf{q}|
\rightarrow 0$ and test this approximation later in a different approach. 

Finite temperature modifications of the vector meson self-energies appearing
in eq.(\ref{ImBarPi}) are calculated using thermal Feynman rules. The explicit
calculations for the $\rho$- and $\phi$-meson can be found in \cite{SW00}.
The thermal spectral
function of the $\omega$-meson is discussed in detail in \cite{SW01}.
The $\rho$ mass is not shifted explicitly in the approach.

For the evaluation of finite baryon density effects which are relevant at SPS conditions,
we use the
results discussed in \cite{KKW2}. There it was shown that in the linear
density approximation, $\Pi_V$ is related to the vector meson - nucleon
scattering amplitude.  In the
following, we assume that the  temperature- and density-dependences of $\Pi_V$
factorize. This amounts
to neglecting contributions from matrix elements such as $\langle \pi N |
\mathcal{T} j_\mu(x) j^\mu(0) | \pi  N \rangle$ describing nucleon-pion
scatterings where the pion comes from the heat bath. 

Our second approach \cite{Amruta} 
(referred to as 'mean field model'
in the following) is based on a different idea:
The method of thermofield dynamics \cite {tfd} is used here to
calculate the state with minimum thermodynamic potential,
at finite temperature and density within the mean field sigma model with a quartic
scalar self interaction. The temperature and density dependent baryon
and sigma masses are calculated  self-consistently.
The medium modification to the masses 
of the $\omega$- and $\rho$-mesons in hot nuclear matter 
including the quantum correction effects are then calculated
in the relativistic random phase approximation. The decay widths for the mesons are
calculated from the imaginary part of the self energy using the
Cutkosky rule.

In a third approach (referred to as '$\pi-\rho$-model' in the following) 
\cite{rho-pi}, we solve truncated Schwinger-Dyson
equations of the $\pi-\rho$ system in a self-consistent resummation scheme, neglecting the effects
of other hadrons. This approach  takes into account the finite
in-medium damping width of the pion which contributes substantially to the broadening
of the $\rho$-meson. As an additional benefit, we can use this model to study the
effects of neglecting the {\bf q} dependence of the spectral function in
chiral model.

\subsubsection{The measured spectrum}

The differential rate of Eq.(\ref{dileptonrates}) is integrated over the space-time history of the collision to
compare the calculated dilepton rates with the CERES/NA45 data \cite{CERES}
taken in Pb-Au collisions at 158 AGeV (corresponding to
a c.m. energy of $\sqrt{s_{NN}} \sim 17$ AGeV).  The CERES experiment is a
fixed-target experiment. In the lab frame, the CERES
detector covers the limited rapidity interval $\eta = 2.1-2.65$, {\em i.e.} $\Delta\eta = 0.55$. 
We integrate the calculated rates over the
transverse momentum $p_T$ and average over $\eta$.
The resulting formula for the space-time- and $p$-integrated dilepton rates is
\begin{equation}
\frac{d^2N}{dM d\eta} =  \frac{2\pi M}{\Delta \eta} \int \limits_{\tau_0}^{\tau_{f}} \tau
d\tau \, 2\pi \int \limits_0^{R(\tau)} r dr 
\negthickspace \negthickspace \negthickspace
\int \limits_{-\eta_{front}(\tau)}^{\eta_{front}(\tau)} 
\negthickspace \negthickspace
\frac{\sinh (\zeta -1)}{(\zeta - 1)}
d\eta_s \int
\limits_0^\infty dp_T \ p_T
 \ \frac{dN(T(\tau),M, \eta(\eta_s),
p_T)}{d^4 x d^4p} \ \text{Acc}(M, \eta, p_T), \label{integratedrates}
\end{equation}
where  the function $\text{Acc}(M, \eta, p_T)$ accounts for the
experimental acceptance cuts specific to the detector. In the CERES
experiment, each electron/positron track is required to have a transverse
momentum $p_T > 0.2$ GeV, to fall into the rapidity interval $2.1 < \eta <
2.65$ in the lab frame and to have a pair opening angle $\Theta_{ee} > 35$
mrad. 
Finally, for comparison with the CERES data, the resulting rate 
is divided by $dN_{ch}/d\eta$, the rapidity density of charged particles.

In addition to the thermal emission of dileptons, we also consider dileptons
from decays of vector mesons $\rho, \omega, \phi$ after the kinetic 
decoupling of the fireball. We calculate their yields using the statistical 
hadronization model outlined in section \ref{S-Hadrochemistry} and use 
their vacuum spectral functions to calculate the resulting dilepton yield.

\subsubsection{Results}

The invariant mass spectrum of dileptons is a Lorentz-invariant quantity and hence independent
of flow and roughly proportional
to the radiating 4-volume in the hadronic phase. 
This explains the negligible differences between the old and the new
analysis in spite of the fact that the refitted evolution scenario exhibits
stronger transverse flow and has a reduced lifetime.
 
\begin{figure}[!htb]
\begin{center}
\epsfig{file=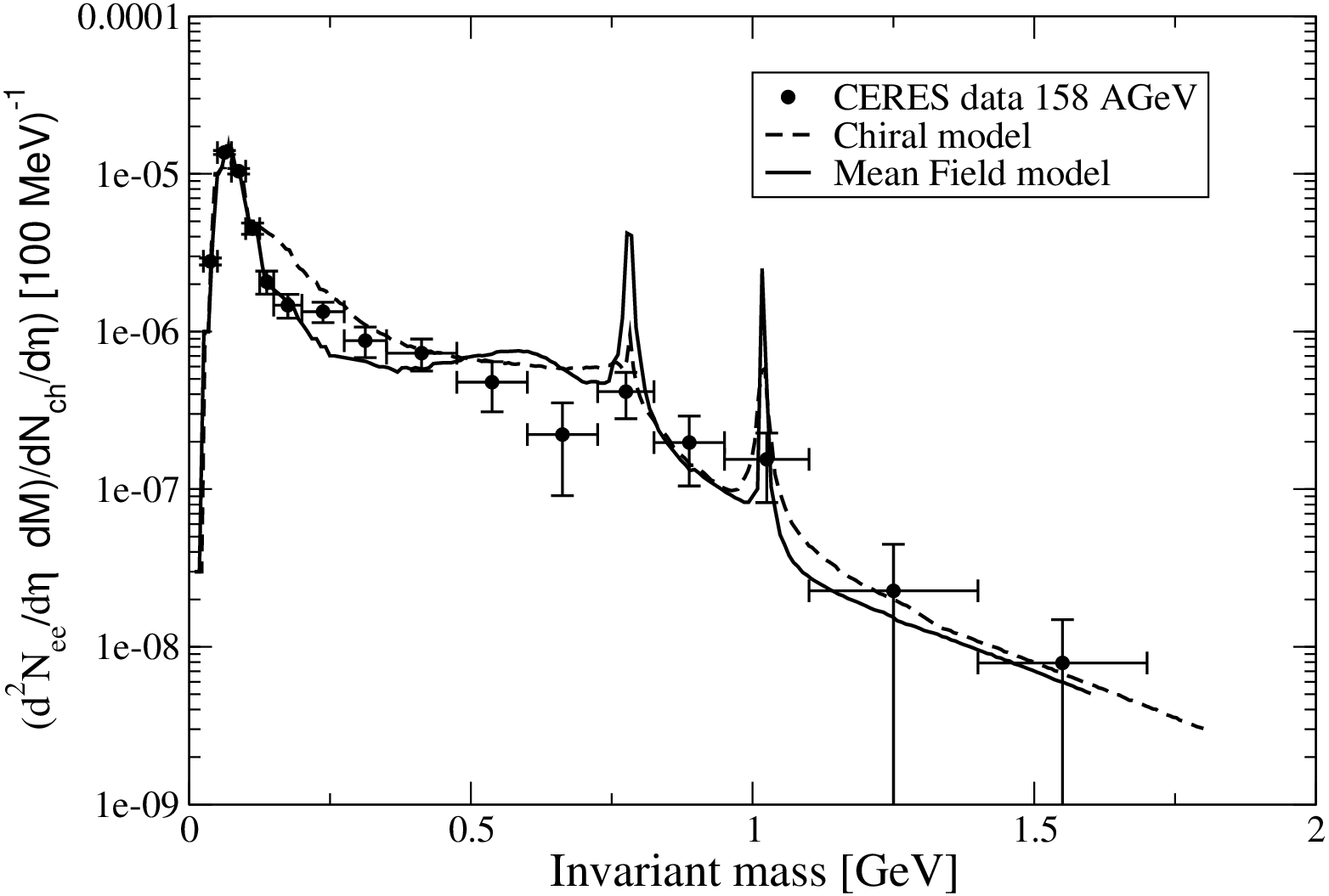, width=6.0cm, angle=-90}
\epsfig{file=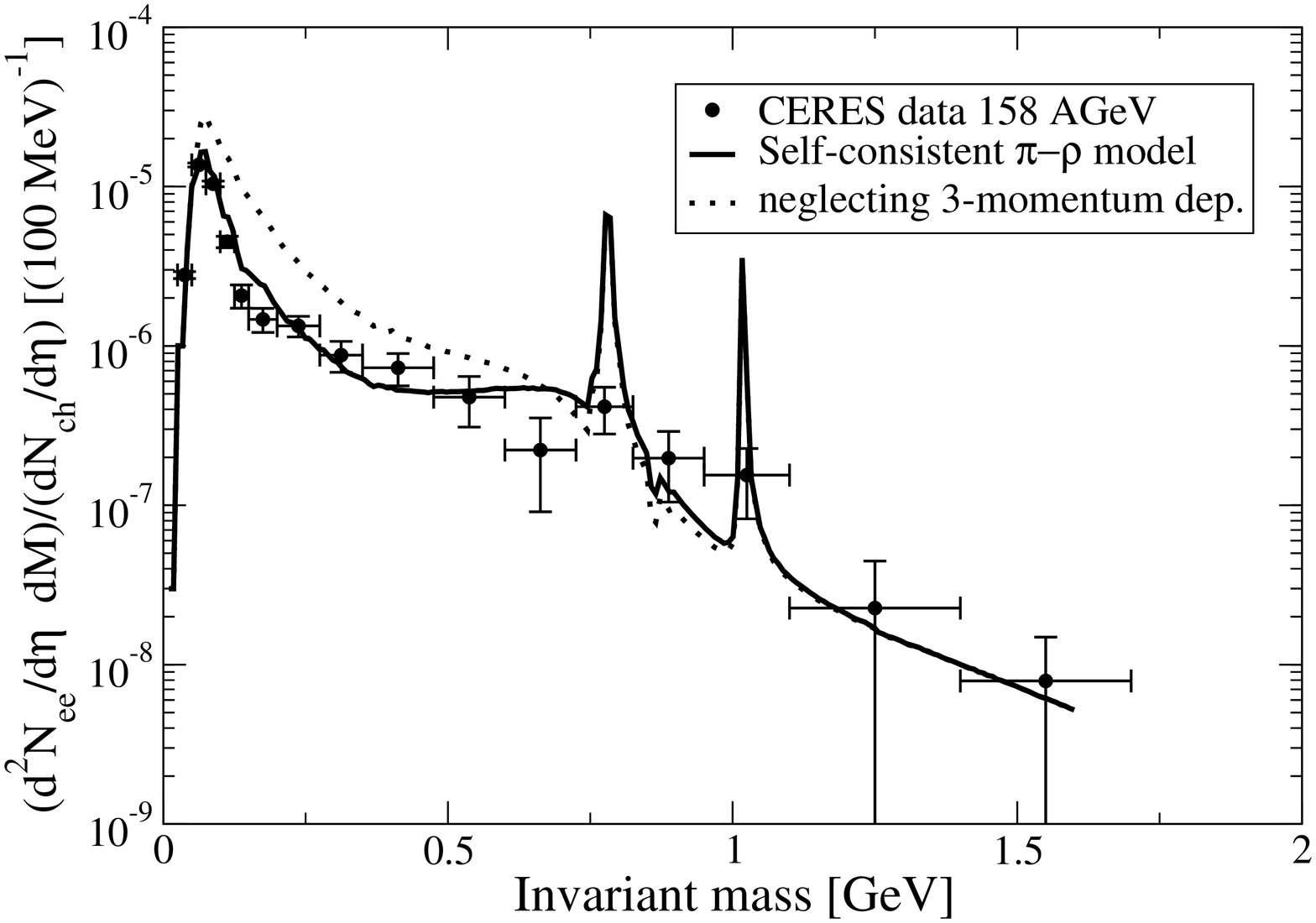, width=6.0cm, angle=-90}
\end{center}
\caption{\label{F-DileptonResult}Left panel: The dilepton spectrum compared with SPS 158 Pb-Au
results \cite{CERES} (dots) using both the Chiral model (dashed) and the Mean Field model (solid)
spectral function. 
Right panel: The dilepton invariant mass spectrum in the self-consistent $\pi-\rho$ model
(solid) and in a calculation neglecting the 3-momentum dependence (dotted). 
Note that although the Mean Field model spectral function does
not contain the $\phi$ meson and the $\pi-\rho$ model contains neither $\omega$ nor $\phi$, 
vacuum decays of these mesons are nevertheless included and lead to distinct
peaks in the spectrum. In both plots we did not
fold our result with the finite energy resolution of the detector, this explains the
apparent disagreement of the sharp $\phi$ and $\omega$ vacuum decay peaks with the data.}
\end{figure}

The calculated spectrum is in fair agreement with the data for all three spectral functions (see
Fig.~\ref{F-DileptonResult}); the mean field model seems superior to the chiral model 
in the description of
the invariant mass region below 300 MeV, however, the general magnitude of
dilepton emission is well reproduced by both spectral functions.
As Fig.~\ref{F-DileptonResult} (right) indicates, a purely pionic medium 
is also able to account for a sufficient broadening of the $\rho$ meson 
if the pion damping width is taken into account in a self-consistent 
resummation scheme instead of using a perturbative expansion. 

Neglecting the momentum dependence in the $\pi-\rho$ model
to overestimates the data in the low invariant mass region. This is the same trend
seen in the chiral model and suggests that including the momentum dependence
there would probably also improve the agreement with the data. 

\begin{figure}[!htb]
\begin{center}
\epsfig{file=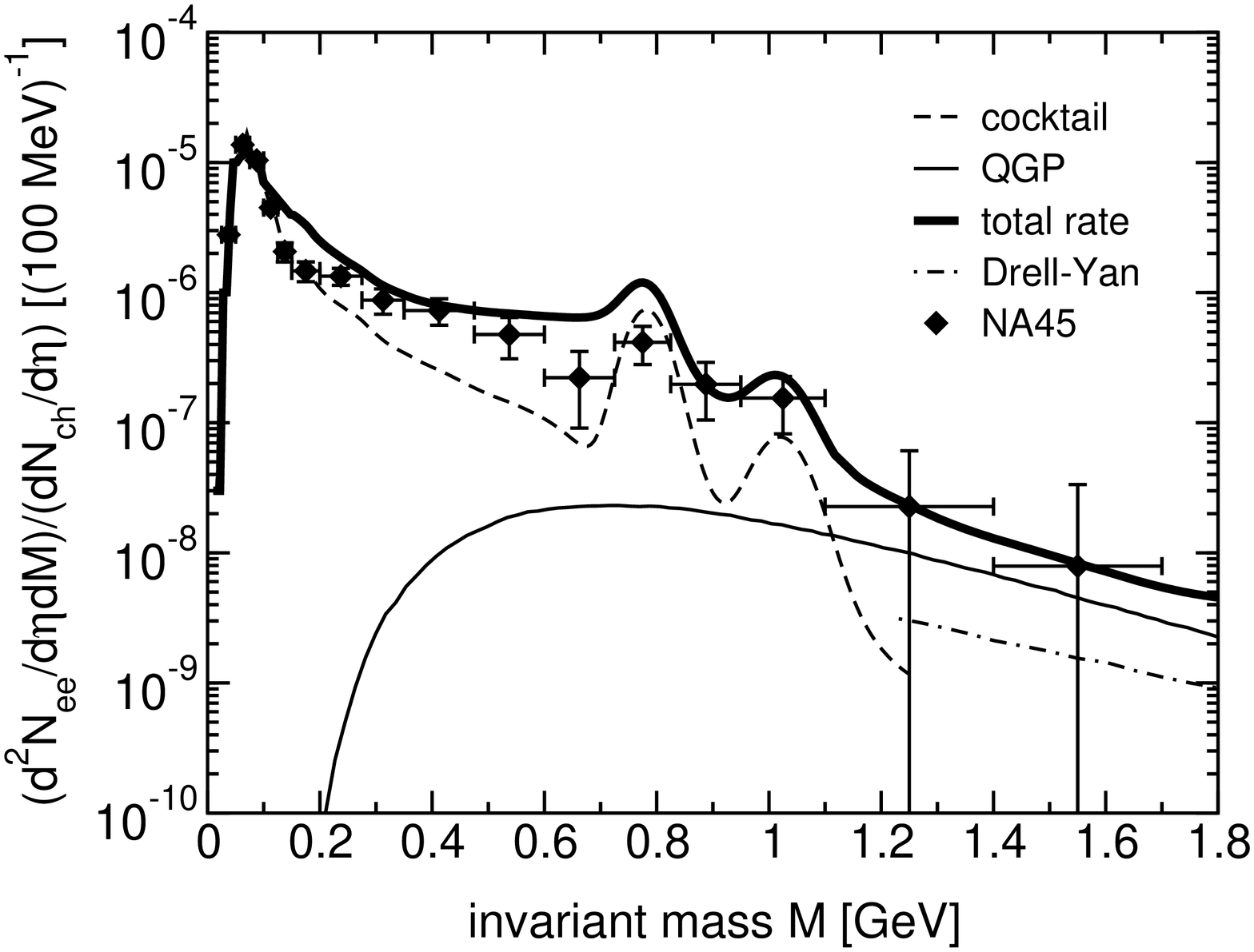, width=6cm, angle=-90}\epsfig{file=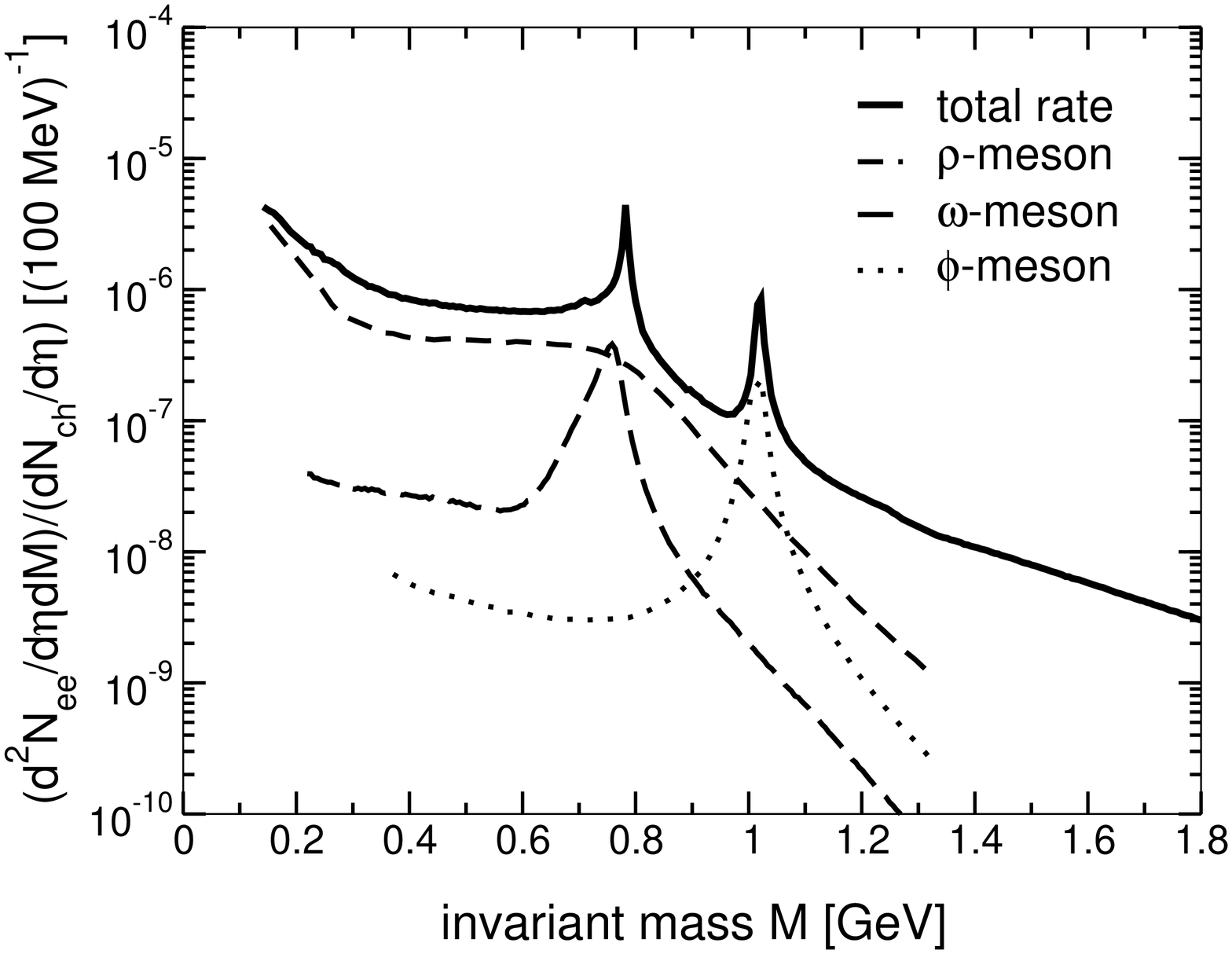, width=6cm, angle=-90}
\end{center}
\caption{\label{F-Decomposed}Left panel: Dilepton emission rate in the Chiral model folded with the
energy resolution of the detector (thick solid) 
compared with SPS 158 AGeV Pb-Au results (solid circles). Indicated are the thermal contribution from the QGP 
(thin solid), contributions from initial hard Drell-Yan processes (dash-dot) and vacuum decays of mesons, the
so-called cocktail (dashed). Right panel: The total thermal emission rate from hadronic degrees of
freedom (solid), decomposed into contributions from $\rho$ (dashed), $\omega$ (long dashed) and
$\phi$ (dotted).}
\end{figure}

To gain additional insight into the underlying physics, we convolute the chiral model
result with the finite energy resolution of the detector and indicate the individual
contributions to the measured rate separately. This is shown in Fig.~\ref{F-Decomposed}.
The decomposition for the Mean Field model is very similar.

The dominant contribution to the complete rate in the invariant
mass region below 1 GeV comes from thermalized hadronic matter, and here by far the
strongest contribution from the $\rho$. For both the Chiral and the Mean Field model
calculation, the inclusion of finite baryon density effects (which manifest themselves in
N-$\rho$ scattering processes) is crucial for achieving agreement with data \cite{Dileptons, Amruta}.
Both calculations take these into account. The success of the $\pi-\rho$ model suggests,
however, that even for vanishing baryon density the $\rho$-meson may be sufficiently
broadened to explain the data.

The QGP contribution is only visible above 1 GeV invariant mass, however here the large errors
do not allow any firm conclusion about its magnitude.
As is apparent from Fig.~\ref{F-Decomposed}, the QGP contribution is unable
to explain the enhancement of the dilepton spectrum below 600 MeV invariant mass.
Thus a dilepton measurement at SPS is dominantly sensitive to the in-medium
mass modifications of vector mesons. While those are interesting problems in their
own right, it implies that the dilepton measurement at CERES does not directly help to 
clarify the question if the quark-gluon plasma has
been formed.

The importance of thermal contributions from the hadronic phase serves as a reminder that
interactions do not cease to be important after the phase transition.
Any type of model assuming kinetic freeze-out at the phase transition would
have to find a different mechanism of producing the required amount
of dilepton radiation below the $\rho$ peak --- clearly, the QGP contribution alone
is insufficient to explain the enhancement. On the other hand, the
fact that the hadronic spectral functions we have investigated lie on the upper and
lower bound of the error bars below 0.5 GeV invariant mass indicates
that the hadronic contribution (dictated by the four-volume of the radiating matter)
is of the correct order of magnitude. If the radiating 4-volume were less than half
of the value in the present model, the calculated hadronic emission spectrum would
not be sufficient to explain the data. This strongly disfavours large freeze-out temperatures
and a short lifetime of the hadronic phase (although for sufficient transverse flow,
a large emitting 4-volume could still be achieved in spite of a small lifetime; the measured
hadron spectra, however, rule out such a combination).

\subsection{Photon emission}

\subsubsection{The emission rate}

For the emission rate of direct photons, we use the complete $O(\alpha_s)$ calculation
\cite{2-2-Kapusta,2-2-Baier,Aurenche1, Aurenche2, Aurenche3,Complete1, Complete2} in the
form of the parameterization provided in \cite{Complete2}.
Using arguments analoguous to the ones in the case of dilepton emission,
we incorporate the effect of the quasiparticle description into this
equation by a factor $C(T)^2$.
For the photon emission rate from hadronic matter, 
we use a parametrization of the
rate from a hot hadronic gas taken from \cite{HHG}.

\subsubsection{The emission spectrum}

\begin{figure}[htb]
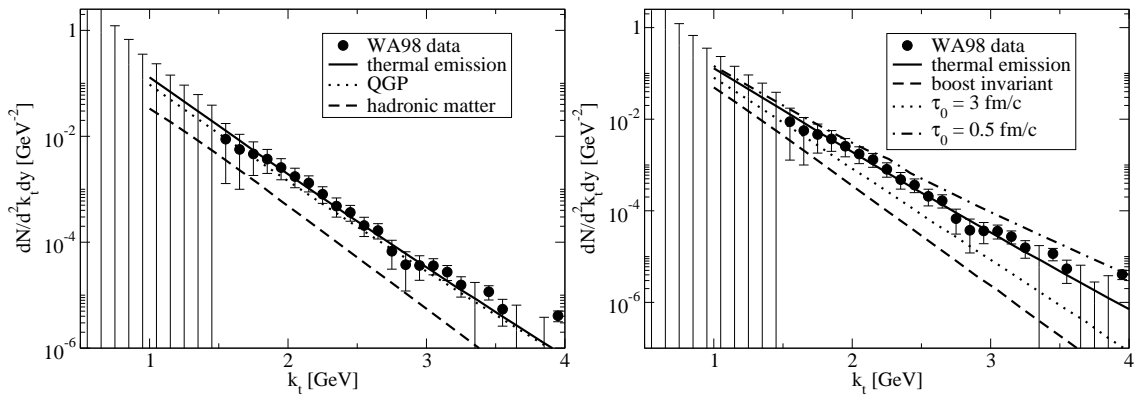

\begin{center}
\vspace{3ex}
\epsfig{file=photon_data.eps, width=7.5cm}\epsfig{file=photon_wrong.eps, width=7.5cm}
\end{center}
\caption{\label{F-Photon-Rates}
Left panel: Thermal photon spectrum for 10\% most central
Pb-Pb collisions at SPS, 158 AGeV Pb-Pb collisions, shown are
calculated rate (total, contribution from QGP and hadronic gas) and
experimental data \cite{PhotonData}.
Right panel: The calculated photon spectrum (solid), the spectrum assuming
boost-invariant longitudinal expansion (dashed) and assuming formation times
0.5 fm/c (dash-dotted) and 3 fm/c (dotted) in comparison with the data \cite{PhotonData}.
}
\end{figure}

The spectrum of emitted photons can be found by folding the rate \cite{Complete2} with the
fireball evolution. In order to account for flow, the energy of a photon emitted
with momentum $k^\mu =(k_t, {\bf k_t}, 0)$ has to be evaluated in the local rest
frame of matter, giving rise to a product $k^\mu u_\mu$ with $u_\mu(\eta_s, r, \tau)$
the local flow profile as decribed in section~\ref{sec_framework}.
For comparison with the measured photon spectrum \cite{PhotonData}, we present the 
differential emission spectrum into the midrapidity slice $y=0$. 
The resulting photon spectrum is shown in
Fig.~\ref{F-Photon-Rates} (left).

The overall agreement with the data is remarkably good.
We observe that the relatively large transverse flow in the hadronic phase in the present scenario
leads to a rather flat contribution with a shape similar to the QGP contribution but smaller. 
For the standard choice $\tau_0 = 1$ fm/c  the calculation seems to underestimate
the data slightly in the high momentum tail, but yields a good description
over the whole range 1.5 GeV $< k_t < $ 3 GeV. The errors allow for a prompt photon contribution
of about half of the magnitude of the thermal yield for the whole momentum range. Note that
the spectrum is almost completely saturated by the QGP contribution --- the hadronic contribution
is about an order of magnitude down.

In order to study the influence of the equilibration time $\tau_0$ and hence the
initial temperature, we present model calculations for the choices $\tau_0 = 0.5$
fm/c and $\tau_0=3$ fm/c in fig.~\ref{F-Photon-Rates} (right). Clearly, a late equilibration
fails to account for the data in the momentum region between 2 and 2.5 fm/c where
prompt photons are not expected to play an important role (cf. the discussion in
\cite{Photons}). On the other hand, an equilibration as early as 0.5 fm/c 
(corresponding to an initial temperature of 370 MeV) produces too many photons.
The direct photon spectrum thus constrains the initial equilibration time at
SPS energies to be close to 1 fm/c.

We may finally use the photon emission rate to test the validity of our
fireball model. Tentatively assuming a standard boost-invariant expansion
pattern with $\eta^\text{front}_0 = \eta^\text{front}_f$, we calculate the resulting
photon emission. This is also shown in Fig.~\ref{F-Photon-Rates}, right panel.
The rapid cooling and the consequently short-lived QGP phase for such a scenario
lead to a severe disagreement with data that cannot be compensated for by any
choice of equilibration time above $0.1$ fm/c. Any initial state photon
emission is bound to fill the larger $k_t$ region of the emission spectrum
and cannot properly compensate for the reduction of radiating four-volume at moderate
temperatures (below $250$ MeV)  in the boost-invariant scenario.
We conclude that longitudinal compression and re-expansion of matter with
all its consequences is
essential for the calculated  thermal photon yield to be compatible with the data
and that the analysis indicates a rather quick equilibration.

\section{Charmonium suppression}

It is often argued that the breakup of bound $c\overline{c}$ states immersed
into hot and dense matter is a good signal for the creation of a QGP.
We investigate this question within the framework of our fireball evolution model.
We refer to $J/\Psi$ and the excited states
$\Psi'$ and $\chi_c$ generically as $\Psi$ in the following. Details of our treatment of the
excited states can be found in \cite{Charm}.
Our main assumption is that charmonia themselves are not thermalized.
This is motived by the observation that the mass scale of charm quarks
is well above the temperature scale even for the initial conditions. 
The thermal production of $c\overline{c}$ pairs is therefore
negligible and the momentum transfer from produced charm quarks
to light constituents of the heat bath is not very efficient.

\begin{figure}[htb]
\begin{center}
\epsfig{file=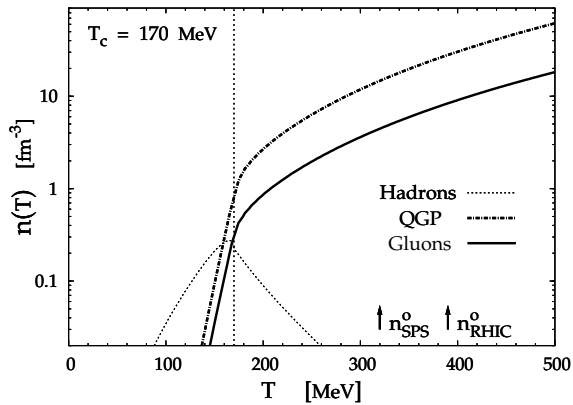, width=5.5cm, angle=-90}
\end{center}
\caption{\label{F-Density}
Density of medium constituents as calculated in the quasiparticle
model \cite{QP1, QP2} as a function of temperature.
}
\end{figure}

Based on this assumption, we will treat the formation and
breakup of bound charmonia as an off-equilibrium process
and use kinetic theory to calculate the evolution of the $\Psi$ density.
This corresponds to the treatment in \cite{CERES-Report}.
We will here limit the discussion to $\Psi$ dissociation in the QGP phase in the
following because the density of scatterers in the medium
as a function of temperature in the hadronic phase is
two orders of magnitude smaller than in the QGP
phase \cite{Charm}. This is illustrated in Fig.~\ref{F-Density} which shows the particle density.
(Fig.~\ref{F-Density})

\subsection{Charmonium production in a nuclear environment}

We start by parametrizing charmonium production in p-p collisions.
For later use we need the $p_T$ spectrum of $\Psi$ at mid-rapidity.
In the following we assume a Gaussian form for the $p_T$-dependent
part, with width parameter $\Lambda = 1$ GeV/c. The rapidity modulation
can be inferred from the relation $d\sigma/dy \sim x_1g(x_1)\,x_2g(x_2)$
where $xg(x) \sim (1-x)^5$ is the gluon distribution in the proton and
$x_{1,2} = (m_\Psi/\sqrt{s})\exp(\pm y)$.
For the overall normalization we use the parametrization for the total 
charmonium production cross section \cite{VOGTREP}
\begin{equation}
\sigma^\Psi_{pp}(s) = 2\, \sigma_0\, (1 - m^\Psi/\sqrt{s})^n\,,
\end{equation}
where
$\sigma_0 = 1.28$ $\mu$b and $n = 12$. 
%

We now consider nuclear effects, starting with the simpler case of 
proton-nucleus ($pA$) collisions. It has been shown that the experimental 
results on charmonium ($\Psi$) production can be explained using \cite{VOGTREP}
\begin{equation}
\sigma^{\psi}_{pA} = \sigma^{\psi}_{pp} \int d^2b
\ T_A(b)\ S^\text{abs}_A(b)\,
\label{protnuc}
\end{equation}
for the total production cross section. The factor
\begin{equation}
S^\text{abs}_A(b) =
\frac{1 - \exp\left[-\sigma^\text{abs}_{\psi N}\,T_A(b)\right]}
{\sigma^\text{abs}_{\psi N}\,T_A(b)}
\label{nucsupp}
\end{equation}
is the survival probability for $\Psi$ to escape the nucleus without being 
dissociated. It includes the effective absorption cross section
$\sigma^\text{abs}_{\psi N}$, a quantity of the order of $3$ mb
for mid-rapidity $\Psi$s as measured at $E_{lab} = 800$ GeV at 
Fermilab, while it amounts to $5-7$ mb for mid-rapidity $\Psi$s as 
measured at $E_{lab} = 158-200$ GeV at the SPS. The absorption cross 
section parametrizes various poorly known effects, with varying importance
depending on the collision energy. Among these effects are the presence of
color degrees of freedom in the dynamics of colliding nucleons, initial state parton
energy loss and coherence length and shadowing effects. A common property
of all of the above is the linear dependence on the path length, at least
to leading order. Using eq.~(\ref{nucsupp}) can therefore be justified,
provided a suitable re-scaling and re-interpretation of 
$\sigma^\text{abs}_{\psi N} \rightarrow \sigma_{\psi N}$
and $S^\text{abs}_{A,B} \rightarrow S^\text{NUC}_{A,B}$ is done.

When considering $\Psi$ production in nucleus-nucleus (AB) collisions,
one can estimate the cross section for a given impact parameter by 
generalizing eq.~(\ref{protnuc}).  Neglecting effects of the
medium produced in the collisions (which will be discussed in the
following) one obtains
\begin{equation}
\frac{dN^{\Psi}_{AB}}{dy\,d^2p_T}(b) = \frac{d\sigma^{\Psi}_{pp}}{dy\,d^2p_T}
\ T_{AB}(b)\ S^\text{NUC}_{AB}(b)\,,
\end{equation}
where nuclear effects are included in the suppression function
\begin{equation}
S^\text{NUC}_{AB}(b) = T_{AB}^{-1}(b)\ \int d^2\vec s
\,\ T_A(s)\, S^\text{NUC}_A(s)\,\ T_B(\tilde s)\, S^\text{NUC}_B(\tilde s).
\end{equation}
We choose $\sigma^\text{abs}_{\psi N} = 5$
mb at the SPS energy $\sqrt{s} = 17.3$ GeV conforming with the 
$pA$ measurement

\subsection{Kinetic description of $\Psi$ breakup}

We describe the suppression of charmonia by using kinetic theory.
Neglecting the possibility of charmonium formation in the medium
for SPS energies, we can derive the expression \cite{Charm}
\begin{equation} 
\frac{dN_\Psi}{d\tau dy}= - \sum_n \langle \langle \sigma_D^n v_\text{rel} \rangle \rangle \rho_n \frac{dN_\Psi}{dy}
\equiv \lambda_D(\tau) \frac{dN_\Psi}{dy}
\label{rateeq}
\end{equation}

for the evolution of the charmonium rapidity density. Here, $\sigma_D^n$ denotes the
dissociation cross section for collisions with medium constituent $n$, $\rho_n$ its density and the sum
runs over all possible constituents.  The double brackets indicate
an average over the momenta of the initial-state particles $\Psi$ and $n$ except the 
$\Psi$ rapidity,  we want to compute the final $\Psi$ rapidity 
distribution at mid-rapidity, not the whole yield. 

The ingredients required to solve this equation are the initial
distribution of charmonium discussed in the previous section, the evolution
of the medium and the functional form of $\sigma_D$.

We use the result of Bhanot and Peskin \cite{P79,BP79} 
\begin{equation}
\sigma_D(\omega) = \frac{2\,\pi}{3} 
\left(\frac{32}{N_c}\right)^2 \!\frac{1}{\mu^{3/2}\!\epsilon^{1/2}} 
\,\frac{\left(\omega/\epsilon -1\right)^{3/2}}{\left(\omega/\epsilon\right)^5}
\label{bpcross}
\end{equation}
for the gluon dissociation process of a heavy quarkonium. It is a function of the
gluon energy $\omega$ in the rest frame of the quarkonium and contains the threshold
energy $\epsilon$ and the mass scale $\mu$, related to the heavy quark mass.
The threshold energy is related to the binding energy $\epsilon_0$ by the condition that 
$s = (p+k)^2 \gsim 4m_c^2$ which implies that $\epsilon=\epsilon_0+\epsilon_0^2/(2 m_\Psi)$.
In the following we assume $\epsilon_0 = 780$ MeV for the binding energy and
$\mu_c = 1.95$ GeV for the mass as in \cite{BP79} to fit the mass values of
the first two levels ($J/\psi$ and $\psi'$) of the charmonium system.

\subsection{Results}

\begin{figure}[t]
\begin{center}
\epsfig{file=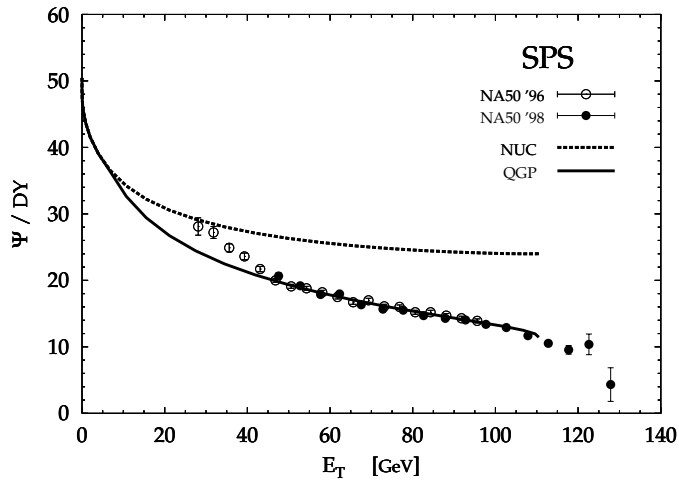,width=6.5cm,angle=-90}
\caption{Result at SPS energy for the ratio $\Psi/DY$ as function of the transverse energy. 
The dashed curve includes only
nuclear effects, while the full line is the complete result including gluon dissociation}
\label{figure:result-sps}
\end{center}
\end{figure}

We combine the building blocks described above to compute the suppression as a function
of impact parameter $b$.
The result is divided
by the number of binary collisions $N_\text{coll}$, which provides the centrality dependence of the Drell-Yan cross 
section. The normalization is fixed at $N_\text{coll}= 53.5$ to match, at large $b$, the experimental
value in p-p collisions and the ratio $\Psi/DY$ as function of $b$ is converted
into a function of the measured transverse energy assuming
$E_T(b) = \epsilon_T\, N_\text{part}(b)$.
The quantity $\epsilon_T = 0.297$, which  represents the amount of produced transverse energy per 
participant, is used in order to describe correctly the total inelastic (minibum-bias) 
cross section as function of centrality. In this way we arrive at the results plotted in 
Fig.~\ref{figure:result-sps}. 

The agreement with data is quite remarkable but  deserves some comments. 
First, the curves 
end at $E_T \approx 110$ GeV, which corresponds to $b=0$. To go beyond this value it is 
necessary to include effects of fluctuations, which are in principle straightforward to address
\cite{fluct}. However, as the analysis in \cite{Capella} suggests, fluctuations do
not alter the results below this value substantially, therefore such an extension of the model
would not yield new information about the scales of the spacetime expansion of the fireball,
the main issue under investigation in this paper.
 
The agreement of our curve with the data down to $E_T = 40$ GeV is remarkable given the
fact that we use rather simple scaling laws \cite{Charm} to
model the impact parameter dependence of the fireball evolution. 
On the other hand, the deviation from the date for $E_T<40 $ GeV is not surprising,
because the
thermal equilibrium assumption is bound to fail for very peripheral collisions.

It is reassuring that the model is able to describe the data over a wide range, however
there are still considerable uncertainties regarding the excited $\Psi$ states,
the magnitude of the 'ordinary' nuclear suppression and the magnitude of the
production cross section for nucleus-nucleus collisions (cf. \cite{Charm}). 
Therefore, we refrain from deriving 
constraints for the model framework from this observable and note only
that the same early evolution which describes the photon data also agrees well
with the charmonium suppression data.

\section{Summary} 

Under the assumption that a thermalized system is created in 158 AGeV Pb-Pb
collisions at SPS, we have analyzed a schematic model based on the thermodynamic
response of matter to a volume expansion. This model provides the link between the
EoS as obtained in lattice QCD and measured observables. We have derived essential
evolution scales by the requirement that the model should reproduce the
measured properties of hadron emission.

Various other sets of observables provide unique opportunities to verify these
essential scales. Discussing statistical hadronization, we find that 
the crucial quantity in our approach is given by $s(T_C)$, the entropy density at
the phase transition. This quantity is an input from lattice QCD
and agrees nicely with the observed hadon ratios.

A calculation of dilepton emission shows that most of the observed dilepton excess above
the 'cocktail' of final state meson decays can be explained by radiation from a thermalized hadron gas.
The main contribution here stems from a strongly broadened $\rho$-meson.  
Thus, we find evidence for ongoing
interactions even below the phase transition temperature. The total amount
of radiating space-time-volume in our model description is in good agreement with the
measured amount of radiation.

The measurement of high-momentum photons gives complementary information: The
large momentum scale makes this observable sensitive to the initial state of the
collision. Again, we find good agreement with the data and 
explicitly demonstrate that a standard boost-invariant expansion does not
yield a satisfactory description. Therefore, initial longitudinal compression and subsequent
expansion, an essential point of our framework, seems to be realized at SPS conditions.

This observation is further strengthened by a calculation of charmonium suppression,
which we also find to be predominantly sensitive to the initial state. Although
uncertainties in the calculation are not small and many simplifications have been made,
the agreement with data is certainly reassuring.

In summary, we have presented a thermal description of relativistic heavy-ion collisions at the CERN SPS
for 158 AGeV Pb-Pb which is highly consistent and leads to a good
description of a large set of different observable quantities. 
While this does not provide a conclusive  proof for the creation of a thermalized system (and hence a
QGP), it constitutes certainly strong evidence for it. It remains to be investigated if the results
can be reproduced in a thermal microscopical transport description (like hydrodynamics) and
if a non-thermal framework is able to describe the experimental results as well or if thermalization is
required by the data.

\section*{Acknowledgements}

I would like to thank B.~M\"{u}ller, S.~A.~Bass, B.~Tomasik, A.~Mishra,
J.~Ruppert, A.~Polleri and R.~A.~Schneider
for interesting discussions, helpful comments and support.
This work was supported by the DOE grant DE-FG02-96ER40945 and a Feodor
Lynen Fellowship of the Alexander von Humboldt Foundation.

\end{document}